\documentclass[lettersize,journal]{IEEEtran}

\usepackage{cite}
\usepackage{amsmath,amssymb,amsfonts}
\usepackage{algorithm}
\usepackage{algpseudocode}
\usepackage{graphicx}
\usepackage{textcomp}
\usepackage{xcolor}
\usepackage{multirow}
\usepackage{epstopdf}
\usepackage{multicol}
\usepackage[symbol]{footmisc}
\usepackage{float}
\usepackage{tikz}
\usepackage{dsfont}
\usepackage{enumitem}
\usepackage{amsthm} 
\newtheorem{problem}{Problem}

\def\BibTeX{{\rm B\kern-.05em{\sc i\kern-.025em b}\kern-.08em
   T\kern-.1667em\lower.7ex\hbox{E}\kern-.125emX}}
\begin{document}

\IEEEoverridecommandlockouts
\IEEEpubid{0000--0000/00\$00.00~\copyright~2025 IEEE}

\title{Multi-User Content Diversity in Wireless Networks}

\author{\IEEEauthorblockN{Belal Korany, Peerapol Tinnakornsrisuphap, Saadallah Kassir, Prashanth Hande, \\Hyun Yong Lee, Thomas Stockhammer, and Hemanth Sampath}\\
\IEEEauthorblockA{\textit{Qualcomm Technologies, Inc.} \\
San Diego, CA, USA \\
\{bkorany,peerapol,skassir,phande,hyunyong,tsto,hsampath\}@qti.qualcomm.com}
}
\markboth{}%
{Multi-User Content Diversity in Wireless Networks}

\maketitle

\begin{abstract}
Immersive applications such as eXtended Reality (XR), cloud gaming, and real-time video streaming are central to the vision of 6G networks. These applications require not only low latency and high data rates, but also consistent and high-quality User Experience (UX). 
Traditional rate allocation and congestion control mechanisms in wireless networks treat users uniformly based on channel conditions, rely only on network-centric Key Performance Indicators (KPIs), and ignore the content diversity, which can lead to inefficient resource utilization and degraded UX.
In this paper, we introduce the concept of \textit{Multi-User Content Diversity}, which recognizes that different users concurrently consume media with varying complexity, and therefore have different bitrate requirements to achieve satisfactory UX. 
We propose multiple different frameworks that exploit multi-user content diversity and lead to overall network-wide gains in terms of UX. For each framework, we demonstrate the required information exchange between Application Servers (ASs), Application Clients (ACs), and the network, and the algorithms that run in each of these components to optimize a network-wide UX-based objective.  Simulation results demonstrate that exploiting multi-user content diversity leads to significant gains in UX capacity, UX fairness, and network utilization, when compared to conventional rate control methods. These findings highlight the potential of content-aware networking as a key enabler for emerging  wireless systems.

\end{abstract}


\section{Introduction}
The emergence of immersive applications such as Extended Reality (XR), cloud gaming, and real-time video streaming is reshaping the landscape of wireless communications. These applications demand not only high throughput and low latency, but also consistent and high-quality user experiences (UX). Current cellular systems (e.g., 4G and 5G) have made significant strides in optimizing network-centric Key Performance Indicators (KPIs) such as data rates and delay. For instance, 3GPP has introduced several enhancements in the 5G system (5GS), e.g., better Quality of-Service (QoS) handling, 5GS information exposure, and application awareness at the network \cite{hande2023extended}. On the QoS front, 5GS introduced the support of PDU-set-based QoS handling, where a PDU-set is a term used to represent a collection of packets that typically carry a single media unit, e.g. a video frame. On the network information exposure front, 5GS adopted Explicit Congestion Notification (ECN) marking for the support of Low Latency, Low Loss, and Scalable Throughput (L4S) traffic \cite{3gpp.23.501}. This means that a 5G network node (e.g., RAN) can mark some IP packets to quickly notify applications of congestion conditions, which helps with rate adaptation at the application layer. As for application awareness, an application may be able to provide the network with PDU-set information (e.g., periodicity, jitter, size,~...) through PDU-set metadata or through standalone assistance information, which helps the network manage its resources \cite{amiri2024application}. 

A key limitation of these enhancements and the existing rate allocation and scheduling mechanisms is their reliance on Channel State Information (CSI) as the only UE feedback to optimize network-centric Key Performance Indicators (KPIs), such as bitrate, over-the-air latency, and packet error rate. These mechanisms assume that users with similar channel conditions and similar application latency requirements should be assigned similar network resources. However, this assumption overlooks a critical dimension of modern media traffic: \textit{the dynamic diversity in content complexity across users}. For instance, a user streaming a static video scene may require far less bitrate than another user viewing a highly dynamic, fast-moving scene to get a good and equivalent user experience—even if both users have identical channel conditions. Allocating resources without accounting for this disparity leads to inefficient utilization and suboptimal User eXperience (UX).

\begin{figure}
\begin{center}
\includegraphics[width=1\linewidth]{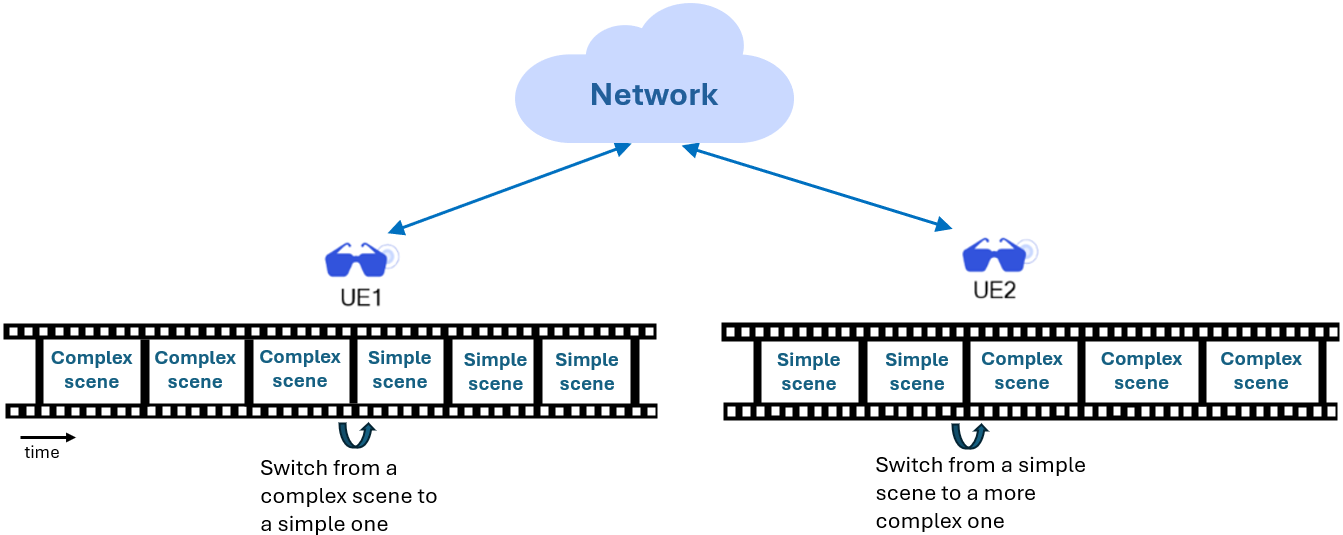}
\caption{Scenario of interest: UEs consuming video traffic with similar channel conditions will receive similar network resources, despite their different video complexities. More efficient management of video traffic can be transformative since it is an elastic traffic that constitutes the majority of mobile traffic volume \cite{ericsson_mobility_2018}.  }
\label{fig:motivation}
\end{center}
\end{figure}

\begin{figure*}
\begin{center}
\includegraphics[width=1\linewidth]{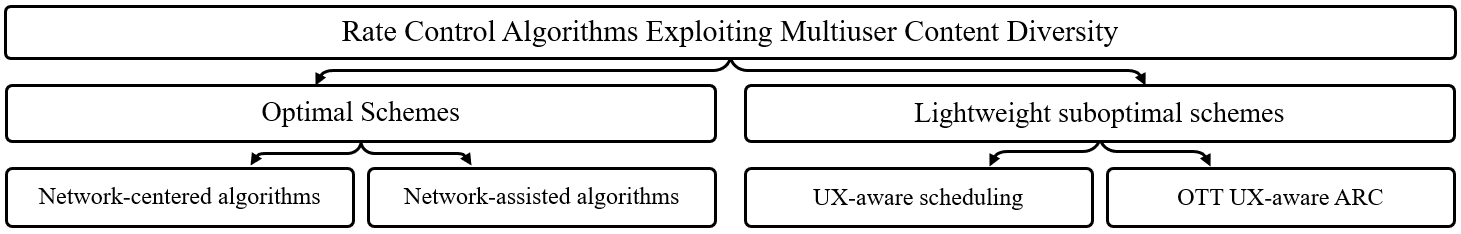}
\caption{Summary of the proposed rate control algorithms to exploit multi-user content diversity.}
\label{fig:alg_summary}
\end{center}
\end{figure*}

\IEEEpubidadjcol
To address this gap, this paper introduces the novel concept of  \textbf{Multi-User Content Diversity}. Generalizing the seminal multi-user channel diversity concept which exploits wireless channel diversity across users to improve system capacity \cite{tse2001multiuser}, this new concept additionally recognizes that in a multi-user wireless network, the content being consumed by each user also varies in complexity (across users and across time, as shown in Fig.~\ref{fig:motivation}), and therefore in its bitrate requirements to achieve a satisfactory UX. For example, for a user live streaming a news broadcast, the picture is mostly static when the news anchor is giving a monologue, requiring relatively low bitrate to achieve a certain quality. When the broadcast switches to a live footage of an event, the spatial and temporal dynamics of the picture typically require higher bitrate to achieve the same quality. As another example, a first-person shooter cloud-gaming application is typically very dynamic, with users emerged in intense battle scenes, fast-paced actions, vibrant graphics, and multi-player interactions. At times, the content may become very simple, e.g., when the player is in a hide-out or pauses the game and navigates the menu.
By explicitly incorporating this content complexity information into the resource allocation process, networks can make more intelligent decisions that better align with the actual needs of each user at any given time.
\IEEEpubidadjcol

The novelty of multi-user content diversity lies in its departure from traditional network-centric paradigms. Rather than allocating network resources to all users based on their channel conditions, we propose various algorithms that incorporate the content complexity information into the rate allocation/control process, and opportunistically free-up network resources to UEs with high-complexity content by decreasing the bitrate of those with low-complexity content, without noticeably impacting their UX.\footnote{Since the UX depends on the content complexity, we use the terms ``UX-awareness'' and ``content-awareness'' interchangeably in the rest of this paper.} As shown in Fig.~\ref{fig:alg_summary}, we cover a wide range of possible rate control algorithms and divide them into two categories: optimal and lightweight suboptimal schemes. In the optimal schemes category, we propose algorithms that optimize some network-wide UX-based objective function. These algorithms may run either on the network side (network-centered), or on the device-side with some network assistance. In the lightweight suboptimal schemes category, we propose heuristic rate control algorithms that still utilize the content complexity information, but are simpler, more lightweight, and require minimal changes to the current devices and/or networks.

The rest of this paper is organized as follows. In Sec.~\ref{sec:related_work}, we discuss the contributions of this paper with respect to the related work. In Sec.~\ref{sec:qoe_model}, we describe the video streaming UX model that will be used in our technical analysis throughout the paper. In Sec.~\ref{sec:optimal_arch}, we describe our proposed optimal rate control schemes, while in Sec.~\ref{sec:suboptimal_arch}, we describe our proposed complexity-reduced rate control schemes.  The simulation results of the proposed algorithms are presented in Sec.~\ref{sec:results}. The signaling and standardization considerations of our proposed schemes are discussed in Sec.~\ref{sec:signaling}. We conclude the paper in Sec.~\ref{sec:conclusion}.

\section{Related Work}
\label{sec:related_work}
At the core of the multi-user content diversity concept is the increased awareness of the network about application characteristics, e.g., the achievable UX. The importance and possible benefits of UX-awareness at the network level has been recognized by few recent papers in the research community \cite{nadas2024qoe,yan2022qoe,slivar2019qoe,liebl2005radio} 
and in standard development organizations, such as 3GPP  \cite{3gpp.22.870}.
In \cite{nadas2024qoe}, the authors recognize the issue that different video streams have different complexities, and that the complexity of a single video stream may vary drastically over time. They propose a resource sharing algorithm that takes this issue into account, albeit, without rigorous validation for the algorithm’s performance. In \cite{yan2022qoe}, the authors propose a Quality of Experience (QoE)-aware resource allocation algorithm for semantic communications, where the QoE model is developed for task-oriented information delivery over the network, which is not suitable for XR traffic. The authors of \cite{slivar2019qoe} also propose a QoE-aware rate allocation framework. However, in their model, each cloud game (or category of games) has one constant time-invariant QoE value, which is not the case for realistic XR traffic. The concept of content-awareness has also been exploited in the context of statistical multiplexing in TV broadcast systems \cite{rezaei2008joint}, where the bitrates of multiple TV channels are dynamically assigned depending on the content complexity. This paper differs from statistical multiplexing in the following aspects: 1) A statistical multiplexer is typically co-located with the different encoders of the TV channels, facilitating the joint optimization of the whole source/transmitter, 2) a TV broadcast statistical multiplexer is AS-based and does not take and application client (AC) feedback/input into account, and 3) in TV broadcasting, all TV channels are multiplexed into a constant-bandwidth channel/medium. In this paper, on the other hand, we consider a wireless system where each UE has a different wireless channel capacity, is connected to an AS which is not necessarily co-located with the ASs of other UEs or with the network entities, UEs can feedback UX information to network.

 In this article, we build upon our previous work \cite{korany2025ux}, in which we propose preliminary versions of the network-centered optimal rate control schemes, and whose importance has been acknowledged by 3GPP, and listed in the 3GPP technical report on 6G use cases and service requirements \cite{3gpp.22.870}. This paper extends our previous work by proposing more categories of algorithms and more simulation results and insights, which demonstrate that exploiting multi-user content diversity leads to significant gains in UX capacity, fairness, and better network utilization, compared to conventional rate control algorithms. These results highlight the transformative potential of content-aware networking in future wireless systems, particularly in the context of 6G and beyond.

\section{User Experience (UX) Model for Real-Time Video Streaming}
\label{sec:qoe_model}
This section introduces and defines a UX model for real-time video streaming for XR services. To maintain the interactiveness of such services, a tight motion-to-photon latency of 20 ms or less is required\footnote{Motion-to-photon latency is the time from a user moving their head until the corresponding display change.} \cite{3gpp.26.928}. The low latency requirement of this traffic necessitates minimal buffering to be implemented on the Application Server (AS) or Application Client (AC), and the media frames to be immediately transmitted from the AS, resulting in a periodic traffic pattern (with some jitter). To avoid queue build-up at the network, the application's bitrate needs to be continuously controlled/adapted to varying channel conditions and network congestion.

To characterize the UX of a real-time video stream, several QoE metrics have been proposed in the literature \cite{min2024perceptual}. These metrics can be broadly classified into two categories: 
\subsubsection{Temporal quality} describing the smoothness of video playback. When a frame is not delivered in time for the device display, the decoder copies the last successfully decoded frame to the display, and the video is said to be in a \textit{stall}. Temporal quality can be measured by the AC, using metrics such as the Maximum Stall Duration (MSD) and stall frequency, which are both functions of the tail of the frame latencies.
\subsubsection{SNR quality} 
\begin{figure}
\begin{center}
\includegraphics[width=1\linewidth]{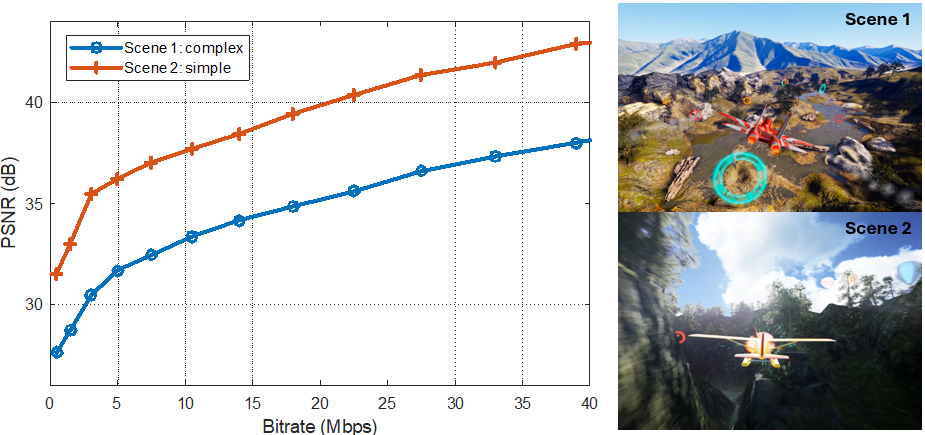}
\caption{Snapshots of different scenes of a cloud game and their RF (PSNR) curves. Scene 1 requires a bitrate of $\sim 19$ Mbps  to achieve a PSNR of 35 dB, while Scene 2 requires only $\sim 3$ Mbps to achieve the same PSNR value.}
\label{fig:scenes}
\end{center}
\end{figure}

describing the quality degradation of the picture due to the artifacts of the compression/encoding process, which is a function of both the scene complexity and the encoding bitrate. SNR quality metrics, such as Peak-Signal-to-Noise-Ratio (PSNR)  \cite{korhonen2012peak} and Video Multimethod Assessment Fusion (VMAF) \cite{rassool2017vmaf}, compare the encoded video frame to the reference non-encoded frame on a pixel-level, block-level, or frame-level. Spatial quality can be measured and/or estimated by the AS during the frame encoding process, and is represented by a Rate-Fidelity (RF) curve, which maps the encoding bitrate to the fidelity (quality) of the frame.  An RF-curve carries crucial information about the content of the video since it depends on the complexity of the video frame(s), where more complex scenes (e.g., ones that are highly dynamic over time) require higher encoding bitrates to achieve the same quality as simple scenes (e.g., ones that are mostly static or slowly moving) encoded with a lower bitrate. For example, Fig.~\ref{fig:scenes} shows the RF curves of two scenes of a cloud game with varying degrees of complexity. Scene 1 (top right) is a complex scene that requires of  bitrate of $\sim 19$ Mbps  to achieve a PSNR of 35 dB, while Scene 2 (bottom right) requires $\sim 3$~Mbps to achieve the same PSNR value. RF curves of complex videos are typically steeper at higher bitrates than those of simpler videos. Moreover, the RF curve of a typical video stream does not remain constant all the time, and changes from one scene to another \cite{nadas2024qoe}. For an interactive XR application, video complexity changes drastically between instances of fast and slow head movement/rotation. For these reasons, an application's time-varying RF-curve can be utilized as a proxy for content information in the rest of the paper.

To unify both quality aspects, a Quality-Bitrate tradeoff curve (QB curve) is established for a specific scene using inputs from both the AS and AC, which maps the encoding bitrate to the overall achieved quality. For simplicity, in this paper, this study targets the temporal quality requirement through the rate allocation algorithm design by ensuring that the video is not in a stall, which we thoroughly discuss in Section \ref{sec:optimal_arch}, and then utilizes only the PSNR RF curve from the AS as the QB curve of the transmitted video. More complex generation of QB curves from both the AS and AC inputs is part of future work. 

\section{Optimal Content-Aware Resource Allocation}
\label{sec:optimal_arch}

In a multi-user scenario, a content-aware network tries to allocate resources/rates to different UEs in order to optimize a network-wide UX-based objective function. Formally, this problem can generally be formulated as an $\alpha$-fair network resource allocation problem \cite{altman2008alphafair}, which provides a unified framework for UX fairness in shared networks:

\begin{equation}\label{eq:ALPHA_GEN_PROBLEM}
\begin{aligned}
\max_{\{R_n\}} \quad & \sum_{n=1}^{N_{\text{UE}}} \frac{\left[ U_n(R_n) \right]^{1-\alpha}}{1-\alpha} \\
\text{s.t.} \quad & \sum_{n=1}^{N_{\text{UE}}} \frac{R_n}{C_n} \le \mu_{\text{target}}, \\
& R_n \ge 0, \quad \forall n,
\end{aligned}
\end{equation}
where $R_n$ is the assigned bitrate to the  $n^\text{th}$ user, $U_n(.)$ is a utility function achieved by the UE, $C_n$ is the channel capacity of the $n^\text{th}$ user (the maximum bitrate they can achieve if assigned all the network resources), $\mu_\text{target}$ is some target network utilization, and $\alpha$ is a fairness parameter. Different choices of $\alpha$ and $U_n$ result in different optimization objectives. In this paper, we focus on two specific cases of special practical interest:

\begin{itemize}[leftmargin=*]
    \item \textbf{UX-capacity maximization (MaxCap):} Following similar definitions of XR Capacity in 3GPP \cite{3gpp.38.838}, we define UX Capacity as the maximum number of UEs per cell, at which at least 90\% of the UEs are satisfied, with a satisfied UE being defined as one whose PSNR is above a threshold $\gamma_n$ more than 95\% of the time, and whose MSD is less than $d_\text{stall}$. For this objective, the network strives to concurrently satisfy as many UEs as possible. This objective is suitable for admission-control networks, where network guarantees satisfaction for admitted UEs. This can be formalized by setting $\alpha=0$ and $U_n(R_n) =  \mathds{1}_{Q_n(R_n) \ge \gamma_n}$ in Eq. \ref{eq:ALPHA_GEN_PROBLEM} to get: 

\begin{problem}\label{prob:general_maxcap}
\begin{equation*}
\begin{aligned}
\max_{\{R_n\}} \quad & \sum_{n=1}^{N_{\text{UE}}} \mathds{1}_{Q_n(R_n) \ge \gamma_n} \\
\text{s.t.} \quad & \sum_{n=1}^{N_{\text{UE}}} \frac{R_n}{C_n} \le \mu_{\text{target}}, \\
& R_n \ge 0, \quad \forall n,
\end{aligned}
\end{equation*}
\end{problem}
where  $Q_n(R_n)$ is the achieved QoE \footnote{Since QoE is a metric of quantifying UX, we use the terms QoE and UX interchangeably in the rest of the paper} of the  $n^\text{th}$ user if assigned a bitrate of $R_n$, $\gamma_n$ is the target QoE, and $\mathds{1}$ is the indicator function. It is worth noting that the content diversity in this formulation is captured in the term $Q_n(R_n)$ which is dependent on the content complexity, as described earlier.

    \item \textbf{UX fairness (MaxMin):} Ensuring equitable UX across all UEs by maximizing the minimum UX. This objective is suitable for best-effort networks, where the network tries to maximize the number of served UEs while maintaining fairness in terms of individual UX. The rate allocation problem can be formalized by setting $\alpha \rightarrow \infty$ and $U_n = Q_n$ in Eq.~\ref{eq:ALPHA_GEN_PROBLEM} to get:
\begin{problem}\label{prob:general_maxmin}
\begin{equation*}
\begin{aligned}
\max_{\{R_n\}}  \min_n \quad &Q_n(R_n)  \\
\text{s.t.} \quad & \sum_{n=1}^{N_{\text{UE}}} \frac{R_n}{C_n} \le \mu_{\text{target}}, \\
& R_n \ge 0, \quad \forall n,
\end{aligned}
\end{equation*}
\end{problem}

\end{itemize}

\begin{figure}
\begin{center}
\includegraphics[width=0.98\linewidth]{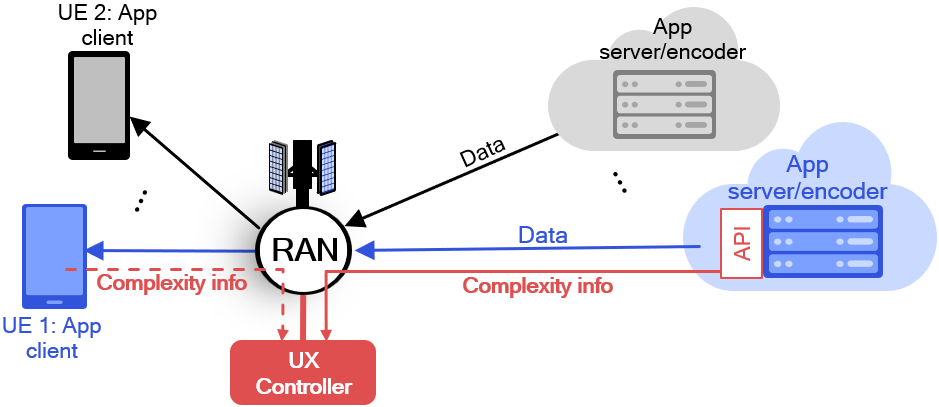}
\caption{Optimal network-centered architecture: A UX Controller logical entity in the network receives content complexity information from application servers/clients and allocates the bitrate accordingly. }
\label{fig:centralized_arch}
\end{center}
\end{figure}

In order to solve these optimization problems, we introduce a new logical entity in the end-to-end communication system, called  \textit{UX rate controller}. The function of the UX-rate controller is to interface with the AS, the AC, and the network to collect the required information to solve the optimization problems, and communicate the allocated rates back to the ASs to update their encoders' bitrates. In terms of where the UX-rate controller resides in the system, we explore two architectural paradigms (which will impact which algorithms are used to solve the rate allocation optimization problems):
\begin{itemize}
    \item \textbf{Network-centered architecture:} The UX rate controller resides within the network (e.g., at the RAN or core network), making centralized decisions on bitrate allocation for all UEs, as shown in Fig.~\ref{fig:centralized_arch}. 
    \item \textbf{Network-assisted architecture:} The UX rate controller is implemented in a distributed fashion, where the network sends assistance information to the UX rate controller instances at the UEs, which then make decentralized bitrate decisions based on local and network feedback, as shown in Fig.~\ref{fig:hybrid_arch}.
\end{itemize}

In the following subsections, we will describe the details of the UX controller and the algorithms it needs to implement to solve the rate allocation problems in both these architectures.

\subsection{Network-centered Architecture}
In the network-centered architecture, the UX rate controller is implemented as a centralized entity within the network infrastructure. It periodically collects QB curves from the application servers and Channel Quality Indicator (CQI) information from the UEs, and computes optimal bitrate allocations for each UE. The centralized nature of this architecture ensures optimality, high performance and coordination, but may require additional signaling overhead and integration with existing network management systems. In the following, we detail the algorithms that the UX rate controller needs to run to solve the optimization Problems \ref{prob:general_maxcap} and \ref{prob:general_maxmin}.

\begin{figure}[!t]

\noindent
\begin{minipage}{1\linewidth}
\begin{algorithm}[H]
\caption{UX Capacity Maximization (MaxCap)}\label{alg:maxcap}
\begin{algorithmic}[1]
\State\textbf{INPUTS (from network):} Number of UEs $N_{\text{UE}}$, Channel capacity per UE $C_n\,\, \forall n$, Target network utilization $\mu_\text{target}$.
\State\textbf{INPUTS (from AS(s)):}  QB function per UE $Q_n(.)\,\, \forall n$, QoE target per UE $\gamma_n\,\, \forall n$

\State\textbf{OUTPUTS:} Allocated bitrate for each UE $R_n$

\State \textbf{Calculate} for each UE, the amount of network resource shares required to meet its target QoE as $ g_n = \frac{ Q_n^{-1}(\gamma_n)}{C_n}$.

\If {$\sum_n g_n \le \mu_\text{target}$}
\State \textbf{Set} $R_n = \left( g_n + \lfloor \frac{\mu_\text{target} - \sum_n g_n}{N_\text{UE}} \rfloor \right) C_n$
\Else
\State Sort UEs in ascending order of $g_n$, with new index $m$
\State Find the maximum number of satisfied UEs as $\max M \text{such that} \sum_{m=1}^M g_m < \mu_\text{target}$
\State \textbf{Set} 
\[ R_m = 
\begin{cases}
 g_mC_m & \quad  \text{for} \,\,m \le M, \\
 \lfloor \frac{\mu_\text{target} - \sum_{m=1}^M g_m}{N_\text{UE} - M} \rfloor C_m &\quad\text{for} \,\, m > M
\end{cases} 
\]

\EndIf
\end{algorithmic}
\end{algorithm}
\end{minipage}
\end{figure}

\subsubsection{UX-Capacity Maximization (MaxCap)}

To maximize the number of UEs with acceptable UX performance, the UX controller solves Problem~\ref{prob:general_maxcap} using the MaxCap algorithm (Algorithm~\ref{alg:maxcap}). The algorithm uses each UE’s QB curve and channel conditions to compute the minimum network share \( g_n \) required to meet its QoE target \( \gamma_n \):

\[
g_n = \frac{Q_n^{-1}(\gamma_n)}{C_n}
\]

UEs are admitted in ascending order of \( g_n \) to maximize the number of satisfied users. If all UEs fit within the target network utilization \( \mu_\text{target} \), remaining resources are evenly redistributed to enhance UX. Otherwise, UEs with the highest resource demands may be excluded.

This approach favors UEs with either high channel capacity or low content complexity. Periodic re-evaluation (e.g., every \( T_\text{period}^\text{qoe} \)) ensures adaptability to changing channel conditions and media characteristics. Congestion and queuing delays are inherently avoided by admitting only UEs that fit within the available capacity.

Note that the UX-capacity maximization algorithm described above solves the UX capacity maximization problem (Problem \ref{prob:general_maxcap}) and can be easily extended to handle other policies of dealing with unsatisfied UEs. For instance, the unsatisfied UEs can be downgraded to meet a lower QoE level (e.g. from excellent QoE to good or acceptable QoE), and the resources can be allocated to them accordingly. Also, Service Level Agreements (SLAs) can play a role in prioritizing the satisfaction of some UEs over others.

\subsubsection{UX fairness (MaxMin)}
For this objective, the UX controller tries to maintain \textit{QoE fairness} among the UEs by maximizing the minimum QoE across the UEs in the cell (Problem \ref{prob:general_maxmin}). Similar to Algorithm \ref{alg:maxcap}, the maxmin fairness algorithm takes the same inputs and starts by calculating the maximum achievable rate $C_n$ of UE~$n$. Then, the algorithm uses the well-known bisection method \cite{oliveira2020enhancement} to search for the rate allocation with which all the UEs in the cell can simultaneously maintain a maximum QoE value in the range $[Q_\text{min}, Q_\text{max}]$, while fitting within the network capacity constraints. The details of the bisection method are summarized in Algorithm \ref{alg:maxmin}, and the readers are referred to \cite{oliveira2020enhancement} for more information about the bisection method.

It is worth noting that, while the concepts of this paper are developed for UEs with real-time media, they are generalizable to cases with mixed traffic. In such cases, each application may model the UX of its underlying traffic and shares its projected QoE as a function of bitrate with the UX rate controller. The UX rate controller may then assign bitrates to the different UEs (using the proposed algorithms) to satisfy their respective UX requirements. Alternatively, each traffic type may be assigned a different priority level by RAN, and the proposed algorithms may then be used to allocate rates for UEs within each traffic priority. Other options for how to deal with mixed traffic scenarios is part of future investigation.

\begin{figure}[!t]

\noindent
\begin{minipage}{1\linewidth}
\begin{algorithm}[H]
\caption{UX Fairness (MaxMin)}\label{alg:maxmin}
\begin{algorithmic}[1]
\State\textbf{INPUTS (from network):} Number of UEs $N_{\text{UE}}$, Channel capacity per UE $C_n\,\, \forall n$, Target network utilization $\mu_\text{target}$.
\State\textbf{INPUTS (from AS(s)):}  QB function per UE $Q_n(.)\,\, \forall n$

\State\textbf{OUTPUTS (to AS(s)):} Allocated bitrate for each UE $R_n$


\State \textbf{Set} arbitrary $Q_\text{max}$ and $Q_\text{min}$

\While {$Q_\text{max} - Q_\text{min} > 0.5 $ dB}
\State Set  $Q_\text{mid}=\frac{Q_\text{max} + Q_\text{min}}{2}$
\State Find the minimum amount of network resource share needed for UE $n$ to maintain  $Q_\text{mid}$ quality,  $g_n = \lceil Q_n^{-1}(Q_\text{mid})/C_n \rceil$

\If { $\sum_n g_n > \mu_\text{target}$ }
\State \textbf{Set} $Q_\text{max} = Q_\text{mid}$.
\ElsIf { $\sum_n g_n <\mu_\text{target}$ }
\State \textbf{Set} $Q_\text{min} = Q_\text{mid}$.
\Else
\State Break
\EndIf
\EndWhile
\State \textbf{Set} the bitrate for UE $n$ as  $R_n = g_n C_n$.
\end{algorithmic}
\end{algorithm}
\end{minipage}
\end{figure}

\begin{figure}
\begin{center}
\includegraphics[width=0.7\linewidth]{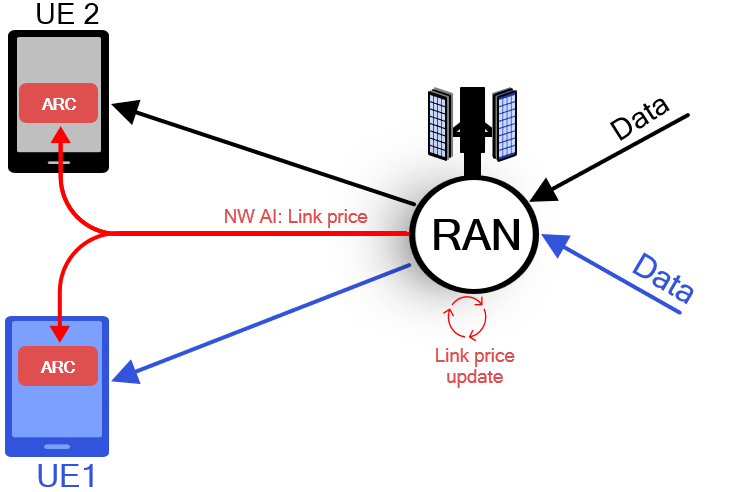}
\caption{Optimal network-assisted architecture: NW shares assistance information related to its current utilization level, and devices utilize this information (along with their own knowledge of content complexity) to optimally adjust their rates locally. }
\label{fig:hybrid_arch}
\end{center}
\end{figure}

\subsection{Network-Assisted Architecture}
In this architecture, the bitrate allocation decisions are made in a distributed fashion at the UEs, but with assistance information from the network. This decentralized approach allows for scalable and flexible implementation while still enabling coordination across UEs.
 Below, we will show how to optimally design the network assistance information, as well as the subsequent device rate allocation updates in order to converge to the same optimal solution for either the MaxCap (Problem \ref{prob:general_maxcap}) or the MaxMin (Problem \ref{prob:general_maxmin}) formulations.

We formulate the problem as a Network Utility Maximization (NUM) problem, which has been extensively studied in the literature \cite{kelly1998rate}, and where the goal is to maximize a network-wide utility function subject to resource constraints. Let $R_n$ denote the bitrate allocated to UE $n$, and let $Q_n(R_n)$ be the QoE achieved by UE $n$ at bitrate $R_n$, as defined by its QB curve. The total network resource usage is constrained by a maximum network utilization  $\mu_{\text{target}}$.
The general NUM problem is formulated as:
\begin{equation}\label{eq:NUM2}
\begin{aligned}
\max_{\{R_n\}} \quad & \sum_{n=1}^{N_{\text{UE}}} U_n(Q_n(R_n)) \\
\text{s.t.} \quad & \sum_{n=1}^{N_{\text{UE}}} \frac{R_n}{C_n} \le \mu_{\text{target}}, \\
& R_n \ge 0, \quad \forall n,
\end{aligned}
\end{equation}

where $U_n(\cdot)$ is a utility function that depends on the optimization objective, and $C_n$ is the maximum achievable bitrate for UE $n$ given its channel conditions.

It has been shown in the literature that this problem can be optimally solved in a distributed manner through an iterative algorithm \cite{wang2006application} (depicted in Fig.~\ref{fig:hybrid_arch}), where the network starts by calculating a \textit{link price} that is related to the current level of network congestion, and broadcasting this price to all the UEs. Each UE then updates its bitrate based on the link price and its local knowledge of content complexity and UX. This process is then periodically repeated until convergence. Mathematically, the network updates the link price $\lambda$ using a gradient ascent method to enforce the resource constraint:
\begin{equation}\label{eq:nw_update}
\lambda(t+1) = \left[ \lambda(t) + \delta \left( \sum_{n=1}^{N_{\text{UE}}} \frac{R_n(t)}{C_n} - \mu_{\text{target}} \right) \right]^+,
\end{equation}
where $\delta$ is a step size parameter and $[\cdot]^+$ denotes projection onto the non-negative orthant. Eq.~\ref{eq:nw_update} shows that the link price keeps decreasing as long as the network has free resources and can accommodate more UEs and/or higher AS bitrates, and starts increasing once the network experiences congestion, i.e., the sum of UE resource shares are higher than the target network utilization. After the link price is shared with the UEs, each UE then solves its own local optimization problem:
\begin{equation}\label{eq:opt_obj}
\max_{R_n \ge 0} \left[ U_n(Q_n(R_n)) - \lambda \frac{R_n}{C_n} \right].
\end{equation}

The definition of the individual utility \( U_n(.) \) and, consequently, the solution to Eq.~\ref{eq:opt_obj} depends on the network-wide objective of the NUM problem. We next detail these definitions and solutions for the MaxCap (Problem \ref{prob:general_maxcap}) and the MaxMin (Problem \ref{prob:general_maxmin}) objectives that were covered in the previous section.

\subsubsection{MaxCap Objective}
For UX-capacity maximization, the utility function $U_n(\cdot)$ is defined as a binary satisfaction indicator:
\[
U_n(Q_n(R_n)) = 
\begin{cases}
1, & \text{if } Q_n(R_n) \ge \gamma_n, \\
0, & \text{otherwise}.
\end{cases}
\]

While this definition renders the UE's local objective function (Eq. \ref{eq:opt_obj}) non-convex, it can be easily shown that the optimal solution is achieved when UEs adjust their bitrate based on the link price as following:
\begin{equation}\label{eq:maxcap_device_update}
R_n(t) = 
\begin{cases}
Q_n^{-1}(\gamma_n), & \text{if }  \frac{\lambda(t) Q_n^{-1}(\gamma_n)}{C_n}< 1, \\
0, & \text{otherwise},
\end{cases}
\end{equation}
where $Q_n^{-1}(\gamma_n)$ is the bitrate needed to satisfy the UE (achieve a UX of $\gamma_n$), and $C_n$ is the maximum achievable rate of the UE. Expectedly, Eq.~\ref{eq:maxcap_device_update} suggests that a UE has a higher chance of being admitted with UX guarantees if the network has abundance of free resources (low $\lambda$), the UE's channel quality is good (high $C_n$), and/or the content of the UE is simple (low $Q_n^{-1}(\gamma_n)$), capturing the content diversity across UEs.

\subsubsection{MaxMin Objective}
For QoE fairness, the objective function of the NUM problem (Eq.~\ref{eq:NUM2}) becomes $\max\min_{R_n \ge 0}Q_n(R_n)$. It has been shown in \cite{wang2006application} that this objective can be achieved by having the UEs adjust their bitrates according to:
\begin{equation}\label{eq:maxmin_device_update}
R_n(t) = Q_n^{-1}\left(\frac{1}{\lambda(t)}\right)
\end{equation}
where $\lambda(t)$ is the current link price of the network, which is updated using the same rule of Eq.~\ref{eq:nw_update}. Inspecting Equations \ref{eq:nw_update} and \ref{eq:maxmin_device_update}, we see that at any point in time, UEs operate at the same UX target of $1/\lambda(t)$, and that, as long as the network has free resources, this target UX keeps increasing (link price decreases) until convergence to maximum possible common UX across UEs.

\section{Lightweight Suboptimal Content-aware Resource Allocation}
\label{sec:suboptimal_arch}

The network-centered and device-assisted algorithms proposed in the previous section are optimal. However, they typically require either a considerable signaling overhead, or standardization of behavior across UEs in their response to network feedback, which may slightly hinder their direct applicability in today's networks. To overcome these additional overheads, in this section, we propose suboptimal content-aware resource allocation strategies that require minimal changes to the existing network infrastructure.
These approaches rely on local decision-making at the UEs or application servers, using heuristics or simplified feedback mechanisms, and while they do not achieve the same level of performance as the optimal algorithms, they still offer significant improvements over traditional rate control techniques, as we shall see in the simulation results section (Section~\ref{sec:results}).

\begin{figure}
\begin{center}
\includegraphics[width=0.9\linewidth]{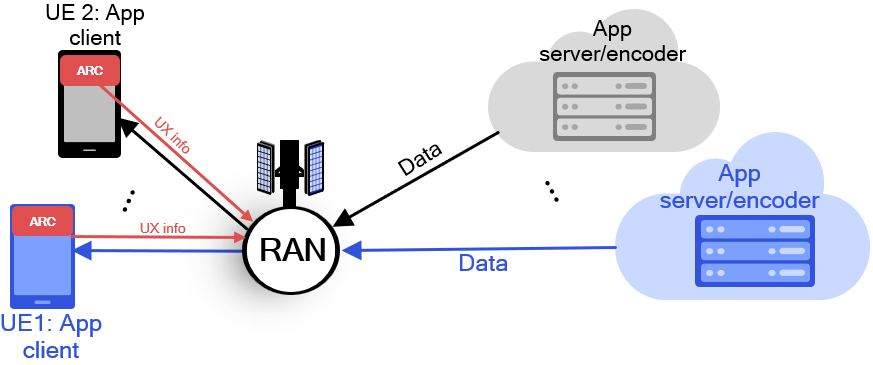}
\caption{Architecture for UX-aware scheduling: UEs share UX information with the network's RAN scheduler. This scheme is to be used in tandem with any traditional ARC algorithm that adjusts the bitrates at the UEs based on the scheduling decisions. }
\label{fig:uxaware_sched_arch}
\end{center}
\end{figure}

\subsection{UX-Aware Scheduler}

Proportional Fair (PF) scheduling is a widely adopted algorithm in wireless networks due to its ability to balance throughput maximization and fairness among users \cite{tse2001multiuser}. The core idea behind PF scheduling is to exploit the time-varying nature of the wireless channels across users and to opportunistically allocate resources to users when their instantaneous channel quality is near its peak. 

Mathematically, the PF scheduler calculates a scheduling metric metric for each user and allocates the network resources to the user with the highest metric. More specifically, the scheduling priority metric of user \( n \) at time \( t \) is given by:

\[
P^\text{PF}_n(t) = \frac{R_n(t)}{\bar{R}_n(t)}
\]

where \( R_n(t) \) is the instantaneous achievable rate of user \( n \) at time \( t \), typically derived from the user's current channel conditions, and \( \bar{R}_n(t) \) is the exponentially averaged throughput of user \( n \) over time.
This metric favors users with good instantaneous channel conditions while penalizing those who have recently received high throughput, thereby promoting fairness over time. PF scheduling is particularly effective in scenarios with fluctuating channel conditions
However, traditional PF scheduling is agnostic to the content being transmitted and its time-varying nature, and the actual user experience (UX). It assumes that all users benefit equally from increased throughput, which is not the case in immersive applications such as XR and cloud gaming, where content complexity varies significantly across users and across time, as described earlier.

To address the limitations of traditional PF scheduling, we propose  \textit{UX-Aware Scheduling}, which exploits both the time-varying nature of the media content as well as the traditional time-varying nature of the wireless channel. Specifically, this approach introduces a new factor into the scheduling metric—\textit{the UX satisfaction factor}, denoted as \( S_n(t) \), which reflects the current quality of experience of user \( n \).
The UX satisfaction factor is a scalar value in the range \( [0, 1] \), where \( S_n(t) \approx 1 \) indicates that the user is highly satisfied, and \( S_n(t) \approx 0 \) indicates very poor UX.
The modified scheduler priority metric then becomes:

\[
P_n^{\text{UX}}(t) = \frac{R_n(t)}{\bar{R}_n(t)} \cdot (1 - S_n(t))
\]

In scenarios of congestion, this formulation opportunistically drives users consuming very simple content (e.g., experiencing extremely high UX  \(S_n(t)\) )  to back-off, without significantly impacting their UX,  and prioritize users experiencing poor UX (i.e., low \( S_n(t) \)). This dynamic adjustment helps the scheduler allocate resources more equitably, improving overall network UX utility.

\begin{figure}
\begin{center}
\includegraphics[width=0.9\linewidth]{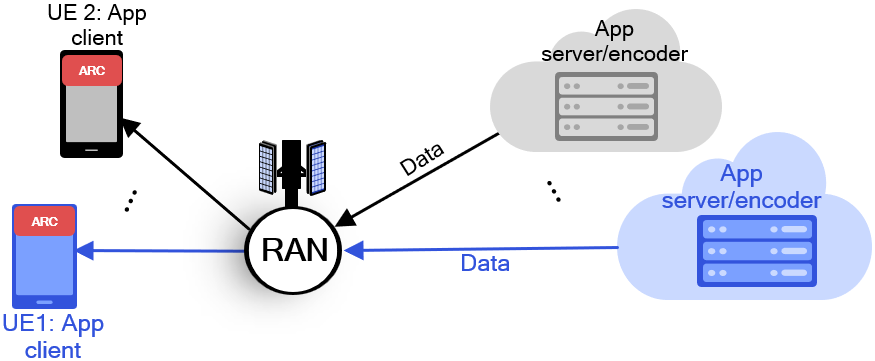}
\caption{Architecture of the OTT UX-aware ARC: Bitrate allocation decisions are taken independently at the UEs with no information exchange between the UEs and the NW or among the UEs. }
\label{fig:distributed_arch}
\end{center}
\end{figure}

Each UE derives its own UX satisfaction factor  \( S_n(t) \) based on local appliation client measurements such as bitrate, PSNR, stall statistics, etc, and reports this factor periodically to the gNB. The exact calculation of  \( S_n(t) \) is left to the application and its definition of user satisfaction\footnote{Policing mechanisms of ensuring applications provide reliable satisfaction measurements are out of scope of this paper.}. A possible calculation for \( S_n(t) \), which this study uses in the simulation results of Sec.~\ref{sec:results}, depends on the allocated bitrate as follows:
\begin{equation}\label{eq:sigmoid}
S_n(t) = \frac{1}{1+e^{-\sigma \left( \bar{R}_n(t) - Q_n^{-1}(\gamma_n) \right)}}
\end{equation}
which approches 1 if the actual achieved bitrate $\bar{R}_n(t)$ is greater than the bitrate required to meet the satisfaction threshold, $Q_n^{-1}(\gamma_n)$, and approaches 0 otherwise. 

It is worth noting that this UX-awareness update of  PF scheduling is not an AS rate control algorithm in and of itself. Rather, it is intended to be applied in conjunction with any traditional adaptive rate control (ARC) or congestion control (CC) algorithm. In congestion scenarios, a user reporting very high satisfaction factor will be deprioritized in scheduling decisions, which will impact its network KPIs (e.g., latency, average throughput, etc.), triggering the application's ARC algorithm to lower the source bitrate, freeing up resources to other users who are reporting bad satisfaction values, which is the required behavior.

\subsection{OTT UX-Aware Rate Control}
\label{sec:ott_ux}

Over-the-top (OTT) rate control algorithms are widely used in real-time media applications, particularly in scenarios where the AS and AC operate independently of the underlying network infrastructure. These algorithms typically rely on end-to-end performance metrics, such as round-trip time (RTT), to infer network congestion and adjust the media bitrate accordingly. While simple and scalable, traditional OTT algorithms inherently focus only on network KPIs and do not account for the actual UX, which depends on both network conditions and content complexity. In the following, this paper describes a basic RTT-based rate control algorithm, and how it can be improved by taking UX measurements into account.

\begin{figure}
\begin{center}
\includegraphics[width=0.7\linewidth]{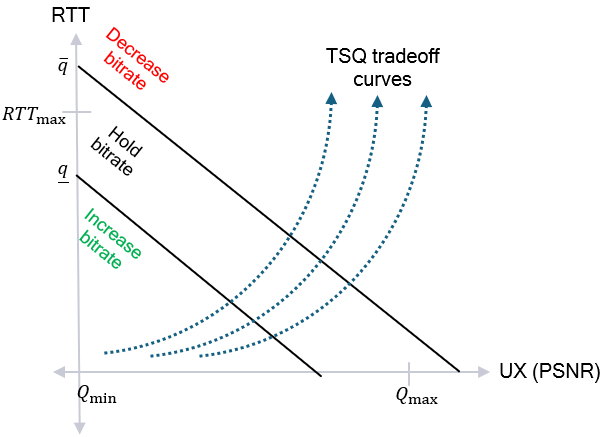}
\caption{Illustration of the OTT UX-aware ARC algorithm: the RTT-PSNR space is divided into three regions, a decrease-bitrate region where either the RTT is too high or the content is too simple (unnecessarily high PSNR), an increase-bitrate region where either RTT is too low or content is too complex, and a hold-bitrate region otherwise. }
\label{fig:distributed_alg}
\end{center}
\end{figure}

\subsubsection{Baseline RTT-based Rate Control}

A common baseline for OTT rate control is the RTT-based algorithm. In this approach, the AC periodically measures the RTT of packets exchanged with the AS and reports the average RTT over a sliding window of duration $T_\text{win}^\text{RTT}$ every $T_\text{period}^\text{RTT}$ milliseconds. The AS then adjusts its encoding bitrate based on the reported RTT as follows:

\begin{itemize}
    \item \textbf{Increase bitrate:} If the average RTT is below a low threshold $\beta_\text{low}^\text{RTT}$, the AS increases the bitrate by a multiplicative factor $\alpha_\text{up}$.
    \item \textbf{Decrease bitrate:} If the average RTT exceeds a high threshold $\beta_\text{high}^\text{RTT}$, the AS decreases the bitrate by a multiplicative factor $\alpha_\text{down}$.
    \item \textbf{Hold bitrate:} If the RTT lies between the two thresholds, the bitrate remains unchanged.
\end{itemize}

This simple control loop allows the AS to react to congestion signals inferred from RTT variations. However, it lacks awareness of the actual quality of the media being delivered, which can lead to suboptimal UX. For instance, a user watching a low-complexity scene may receive unnecessarily high bitrate if they have a good channel (low RTT), while another user having a poor UX due to high-complexity content may not receive sufficient bitrate to maintain acceptable quality.

\subsubsection{Proposed UX-Aware RTT-based Rate Control}

To address these limitations, this paper proposes an enhanced OTT rate control algorithm that incorporates UX-awareness by jointly considering RTT and video quality metrics, specifically PSNR. In the RTT-PSNR 2D space shown in Fig.~\ref{fig:distributed_alg}, the dotted lines represent the temporal-spatial quality (TSQ) tradeoff curves. For a fixed network and content complexity scenario, increasing the source bitrate moves the operating point along one of these curves, enhancing the spatial quality (PSNR), while negatively impacting the frame delays (RTT) due to the larger frame sizes, which could possibly introduce stalls and hurt the temporal quality aspect. Different network conditions and/or content complexity changes the operating tradeoff curve, but still exhibits the same characteristics. 

The key idea of our proposed UX-aware rate control algorithm is to divide the RTT-PSNR space using two lines/functions of negative slope (as shown in the figure), guaranteeing intersection with the TSQ tradeoff curves, where the intersection points represent acceptable fair tradeoff between PSNR and RTT. This divides the space into three regions—\textit{increase bitrate}, \textit{decrease bitrate}, and \textit{hold bitrate}. When the AC reports a $(\text{RTT}, \text{PSNR})$ measurement tuple to the AS every $T_\text{period}^\text{RTT}$ milliseconds, the AS uses the reported RTT and PSNR to determine the appropriate bitrate adjustment action based on the region in which the reported tuple falls, as follows:

\[
R(t+1) =
\begin{cases}
\alpha_\text{up} \cdot R(t), & \text{if } \frac{RTT}{RTT_\text{max}} + \frac{UX - Q_\text{min}}{Q_\text{max} - Q_\text{min}} < \underline{q} \\\\
\alpha_\text{down} \cdot R(t), & \text{if }\frac{RTT}{RTT_\text{max}} + \frac{UX - Q_\text{min}}{Q_\text{max} - Q_\text{min}} >\bar{q} \\\\
R(t), & \text{otherwise}
\end{cases}
\]

\begin{table}
\caption{Simulation Parameters}
\label{tab:sim_params}

\begin{center}
\begin{tabular}{|| c |  c | c ||} 
 \hline\hline
 \multirow{2}{*}{\textbf{Parameter}} & \multicolumn{2}{|c||}{\textbf{Value}} \\ 
\cline{2-3}
  & InH Channel & UMa Channel \\
 \hline\hline
  \multicolumn{3}{||c||}{\textbf{Network parameters}} \\
\hline
Carrier Frequency & 3.5 GHz & 4.7 GHz \\ 
\hline
 ISD & 20 m& 200 m  \\
\hline
 \# of gNBs & 12 & 7  \\
\hline
 \# of cells per gNB & 1 & 3  \\
\hline
Max gNB power & 23 dBm & 44 dBm  \\
\hline
Bandwidth & \multicolumn{2}{|c||}{100 MHz } \\  
 \hline
 SCS  & \multicolumn{2}{|c||}{30 KHz} \\  
\hline
 Noise Figure  & \multicolumn{2}{|c||}{gNB: 5 dB, UE: 9 dB} \\  
\hline
 Scheduler & \multicolumn{2}{|c||}{Proportional Fair} \\  
\hline
Backhaul delay & \multicolumn{2}{|c||}{1 ms} \\ 
\hline
 Target BLER  & \multicolumn{2}{|c||}{10\%} \\  
\hline
 Number of RBGs per slot  & \multicolumn{2}{|c||}{4} \\  
\hline
 Slot pattern  & \multicolumn{2}{|c||}{DDDSU} \\  
 \hline\hline
  \multicolumn{3}{||c||}{\textbf{Source parameters}} \\
\hline
Allowable source bitrates &  \multicolumn{2}{|c||}{1-50 Mbps} \\ 
\hline
Source fps &  \multicolumn{2}{|c||}{60} \\ 
 \hline
Average scene duration &  \multicolumn{2}{|c||}{3.5 seconds} \\ 
 \hline
Encoding delay  &  \multicolumn{2}{|c||}{1 ms} \\ 
\hline
Decoding delay  &  \multicolumn{2}{|c||}{1 ms} \\ 
\hline\hline

   \multicolumn{3}{||c||}{ \textbf{Baseline RTT-based rate control algorithm parameters}} \\
\hline
 $T_\text{period}^\text{RTT}$ &  \multicolumn{2}{|c||}{50 ms} \\ 
 \hline
$T_\text{win}^\text{RTT}$ &  \multicolumn{2}{|c||}{100 ms} \\ 
\hline
$\alpha_\text{up},\, \alpha_\text{down}$ &  \multicolumn{2}{|c||}{1.1, 0.9} \\ 
 \hline
$\beta_\text{low}^\text{RTT},\,\beta_\text{high}^\text{RTT}$ &  \multicolumn{2}{|c||}{8 ms, 10 ms} \\ 
\hline\hline
 \multicolumn{3}{||c||}{\textbf{Prague CC framework parameters}} \\
\hline
 $\beta_\text{low}^\text{L4S}$ &  \multicolumn{2}{|c||}{4 ms} \\ 
\hline
$\beta_\text{high}^\text{L4S}$ &  \multicolumn{2}{|c||}{17 ms} \\ 
\hline\hline
  \multicolumn{3}{||c||}{\textbf{Optimal Content-Aware Rate Allocation}} \\
\hline
$\mu_\text{target}$&  \multicolumn{2}{|c||}{0.9} \\ 
\hline
 $T_\text{period}^\text{qoe}$ &  \multicolumn{2}{|c||}{33 ms} \\ 
\hline
 QoE target $\gamma_n$ &  \multicolumn{2}{|c||}{35 dB PSNR $\,\, \forall n$} \\ 
\hline
Max stall duration ($d_\text{stall}$) &  \multicolumn{2}{|c||}{100 ms} \\ 
\hline
$Q_\text{min}, Q_\text{max}$ (maxmin alg.) &  \multicolumn{2}{|c||}{30 dB, 40 dB PSNR} \\
\hline
Link price step $\delta$ &  \multicolumn{2}{|c||}{0.01}\\
\hline\hline
 \multicolumn{3}{||c||}{\textbf{UX-Aware PF Scheduler}} \\
\hline
Temperature Parameter $\sigma$ (Eq.~\ref{eq:sigmoid}) &  \multicolumn{2}{|c||}{0.5}\\
\hline\hline
 \multicolumn{3}{||c||}{\textbf{OTT UX-Aware Rate Control}} \\
\hline
$RTT_\text{max}$  &  \multicolumn{2}{|c||}{17 ms}\\
\hline
$Q_\text{min}, Q_\text{max}$&  \multicolumn{2}{|c||}{30 dB, 40 dB PSNR} \\
\hline
$\bar{q}, \underline{q}$ &  \multicolumn{2}{|c||}{1.1, 0.9} \\
\hline
$\alpha_\text{up}, \alpha_\text{down}$ &  \multicolumn{2}{|c||}{1.1, 0.9} \\
\hline\hline
\end{tabular}
\end{center}

\end{table}

\begin{figure*}
\begin{centering}
\includegraphics[width=0.9\linewidth]{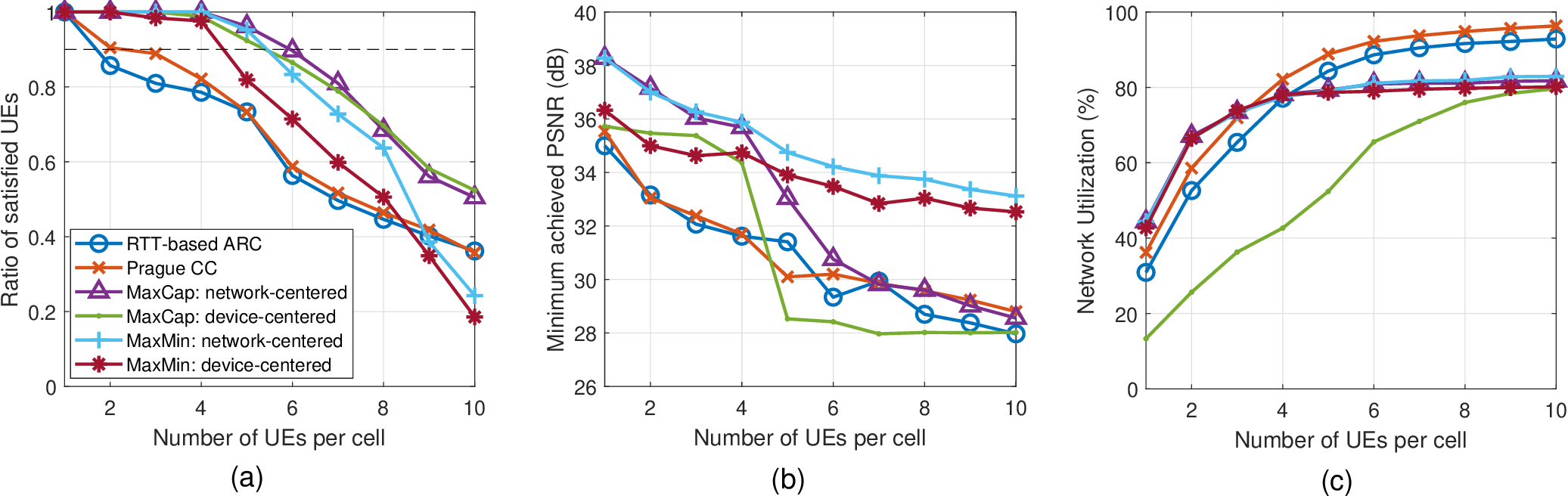}
\caption{Performance evaluation of the optimal schemes: (a) Ratio of satisfied UEs as a function of the number of UEs per cell, showing considerable gains compared to baseline algorithms.  (b) Minimum achieved quality (PSNR) in the cell as a function of the number of UEs per cell. (c) Network utilization (number of occupied RBs divided by the total number of available RBs) as a function of UEs per cell. The proposed algorithms outperform the baseline algorithms while achieving similar or lower network utilization.}
\label{fig:main_plots_optimal}
\end{centering}

\end{figure*}

The proposed algorithm is lightweight and can be implemented with minimal changes to existing OTT streaming frameworks. PSNR estimation can be performed at the decoder using reference-free techniques \cite{zheng2024video} or approximated using encoder-side RF curves \cite{tu2005rate}. The feedback overhead is minimal, as RTT and PSNR can be quantized and piggybacked on existing control messages. Nevertheless, the lack of coordination between the UEs results in a suboptimal performance (as will be verified in the results section). A UE experiencing very high quality may reduce its bitrate to free up resources to other UEs in the system, even if the network is not experiencing congestion. It is also worth noting that this framework assumes all the UEs in the network to utilize the same thresholds/decision rules, which requires standardization of UE behavior.

\section{Simulation Results}
\label{sec:results}
In this section, this paper presents the performance evaluation results for our proposed content-aware rate allocation algorithms. We first list our simulation parameters, describe the baseline algorithms against which this study compares our proposed algorithms, and finally show the simulation results. 

\subsection{Simulation Parameters}
Table \ref{tab:sim_params} lists the simulation parameters for our performance evaluation platform. We first generate SINR traces for the UEs according to the 3GPP Indoor Hotspot (InH) and Urban Macro (UMa) channel models \cite{3gpp.38.901}.
These result in a total of 33 cells (12 InH cells and 21 UMa cells) where the number of UEs per cell is swept from 1 to 10 UEs. The SINR traces are then used to simulate Over-The-Air (OTA) real-time 60-fps video transmission to the UEs. Each UE is sent a gaming video comprising of different scenes that vary in complexity, two of which are shown in Fig. \ref{fig:scenes}. The moments of switching between the scenes are randomized across the UEs.

We compare the performance of our proposed content-aware rate allocation algorithms to two conventional rate control algorithms, which can be broadly classified into Over-The-Top (OTT) algorithms, and network-assisted algorithms.

\subsubsection{RTT-based Rate Control} 
This is the same baseline RTT-based rate control algorithm described in Sec.~\ref{sec:ott_ux}, in which the AC at the UE sends a feedback report to the AS with the average measured RTT within a window of duration $T_\text{win}^\text{RTT}$. The AS increases the bitrate if the average RTT is less then $\beta_\text{low}^\text{RTT}$, decreases the bitrate if the average RTT is above $\beta_\text{high}^\text{RTT}$, and holds the bitrate unchanged otherwise. Similar RTT-based rate control algorithms have been proposed in the literature \cite{maura2024experimenting}.

\subsubsection{Prague Congestion Control} Low Latency, Low Loss, and Scalable Throughput (L4S) is one example of network-assisted frameworks which is standardized by IETF in RFC~9330 \cite{rfc9330}. A network node (e.g., RAN) marks the IP packets using the Explicit Congestion Notification (ECN) field in the IP packet header, with a marking probability that is an increasing function of the queueing delay experienced at the node. The marking policy in our implementation is to have a zero marking probability for queueing delay $\le \beta_\text{low}^{L4S}$ ms, a 100\% marking probability for delays $\ge \beta_\text{high}^{L4S}$ ms, and linear in between. Finally, an L4S-compliant end-to-end rate adaptation algorithm utilizes these markings to adjust the source bitrate. Prague Congestion Control \cite{briscoe2019implementing} is one such rate adaptation algorithm that this study utilizes as a baseline for this study, and is characterized by: 1) additive bitrate increase for every unmarked packet, 2) multiplicative decrease (once per RTT) for marked packets by a factor of $(1-m_\text{ecn}/2)$, where $m_\text{ecn}$ is the fraction of recently marked packets, and 3) multiplicative decrease upon packet loss by a factor of 1/2.

The parameter values of the baseline algorithms used in our study are provided in Table \ref{tab:sim_params}.

\begin{figure*}
\begin{centering}
\includegraphics[width=0.9\linewidth]{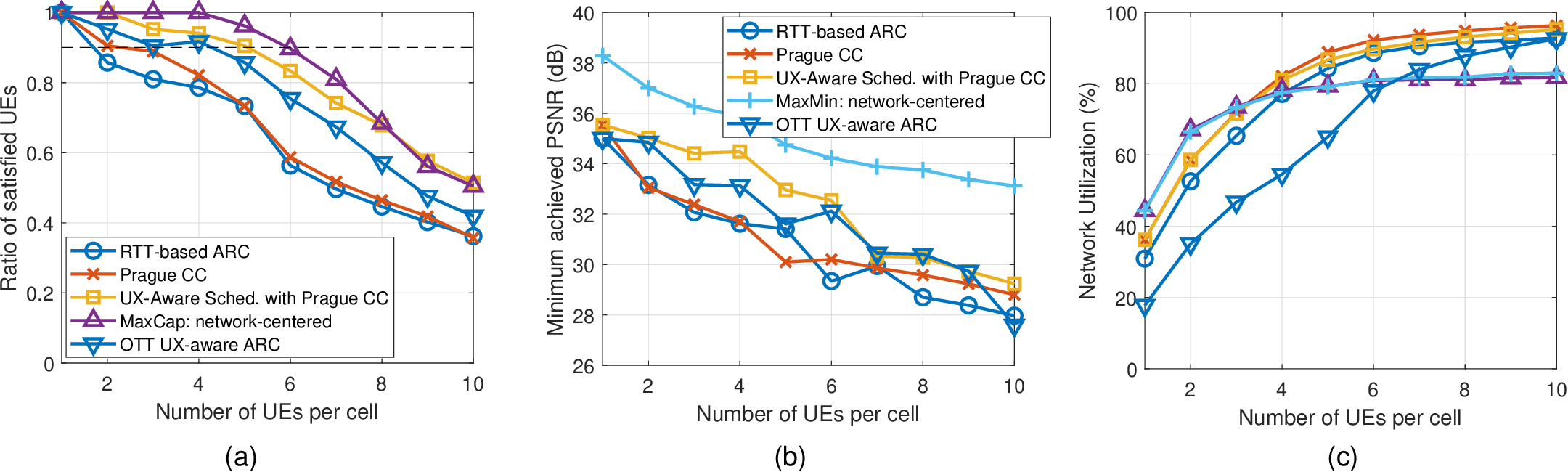}

\caption{Performance evaluation of the suboptimal schemes: (a) Ratio of satisfied UEs as a function of the number of UEs per cell, showing considerable gains compared to baseline algorithms. Performance of the optimal MaxCap algorithm is plotted for comparison.  (b) Minimum achieved quality (PSNR) in the cell as a function of the number of UEs per cell. Performance of the optimal MaxMin algorithm is plotted for comparison. (c) Network utilization (number of occupied RBs divided by the total number of available RBs) as a function of UEs per cell. The proposed algorithms outperform the baseline algorithms while achieving similar or lower network utilization.}
\label{fig:main_plots_suboptimal}
\end{centering}

\end{figure*}

\subsection{Simulation Results}
\subsubsection{Optimal schemes}
Fig. \ref{fig:main_plots_optimal} shows the simulation results for the optimal schemes explained in Section~\ref{sec:optimal_arch}.  Fig. \ref{fig:main_plots_optimal} (a) shows the UE satisfaction rate as a function of the number of UEs per cell. A satisfied UE is one whose PSNR is above a threshold $\gamma_n$ more than 95\% of the time, and whose maximum stall duration is less than $d_\text{stall}$. At 6 UEs per cell, it can be seen that the ratio of satisfied UEs is $\sim$89\% with the network-centered MaxCap algorithm, and $\sim$83\% with the network-centered MaxMin algorithm, compared to $\sim$58\% satisfaction ratio with Prague congestion control, and $\sim$56\% satisfaction ratio with RTT-based rate control. Following similar definitions of XR capacity in 3GPP \cite{3gpp.38.835}, we define UX capacity as the maximum number of UEs per cell, where at least 90\% of the UEs in that cell are satisfied.  It can be seen from Fig. \ref{fig:main_plots_optimal} (a) that the UX capacity of the proposed network-centered algorithms is~5~UEs, while that of Prague congestion control is~3~UEs and RTT-based rate control is~1~UE, showing that content-aware rate control provides a $\sim$67\% UX capacity gain when compared to conventional rate control algorithms.

Fig. \ref{fig:main_plots_optimal} (b) shows the minimum achieved PSNR by the different algorithms as a function of the number of UEs per cell. By definition, the network-centered MaxMin algorithm outperforms all the other algorithms in terms of this KPI. In contrast, for high number of UEs, if the MaxCap algorithm cannot satisfy all the UEs, the  ``unadmitted" UEs may experience very bad quality, lowering the minimum achieved PSNR significantly. It is worth noting that the gains achieved by the network-centered algorithms do not come at the expense of higher network utilization (as shown in Fig.~\ref{fig:main_plots_optimal} (c)). 

While it has been theoretically proven (in Sec.~\ref{sec:optimal_arch}) that the network-assisted algorithms converge to the same solution as the network-centered algorithm, Fig.~\ref{fig:main_plots_optimal} shows a slightly worse performance for the network-assisted algorithms. The reason for this deterioration is the reactive nature of the network-assisted algorithms, where the network needs to first experience a bad congestion event before sending an updated link price to the UEs, pushing them for the optimal rate control solution, slightly degrading the performance in this short transitory period.

\begin{table*}[ht]
\centering
\caption{Information Exchange Between Applications and Network for Proposed Schemes}
\label{tab:info_exchange}
\begin{tabular}{|l|c|c|c|c|}
\hline
\textbf{Direction} & \textbf{Network-Centered Optimal} & \textbf{Network-Assisted Optimal} & \textbf{UX-Aware PF Scheduler} & \textbf{OTT UX-Aware ARC}  \\
\hline
AS/AC $\rightarrow$ RAN & QB curve, target QoE & None & UX Satisfaction factor & None  \\
\hline
RAN $\rightarrow$ UE/AC & Assigned bitrate & Link price broadcast & None & None  \\
\hline
\end{tabular}
\end{table*}

\subsubsection{Lightweight suboptimal schemes}

Fig. \ref{fig:main_plots_suboptimal} shows the performance of the suboptimal schemes proposed in Sec.~\ref{sec:suboptimal_arch}. Fig.~\ref{fig:main_plots_suboptimal} (a) shows the ratio of satisfied UEs as a function of the number of UEs per cell, depicting that the UX-aware PF scheduler achieves a UX capacity of 5~UEs, a gain of 66\% over the conventional PF scheduler Prague Congestion control. The optimal network-centered MaxCap algorithm (Algorithm \ref{alg:maxcap}) performance is plotted for comparison. At capacity (5 UEs per cell), the UX-aware PF scheduler achieves a 17\% increase in overall UE satisfaction ratio over the conventional content-unaware algorithms. In terms of fairness (minimum achieved PSNR), Fig.~\ref{fig:main_plots_suboptimal} (b) shows that the UX-aware PF scheduler achieves a better UX fairness than the baseline algorithms. These resuls show that the UX-aware PF scheduler strikes a great balance between performance and complexity, and can be realized by minor changes to the conventional PF scheduler with lightweight signaling overhead. The OTT UX-aware algorithm is also shown to outperform the baseline algorithms (25\% increase in UX capacity and 13\% increase in UE satisfaction at 5 UEs per cell). However, underutilizes the network for small number of UEs per cell, since it drives UEs with simple content to lower their bitrate even if the network is not experiencing a congestion and the resources are not needed for other UEs. This is a result of the fact that, in this scheme, there is no coordination among the UEs or between the UEs and the network.

\section{Signaling Aspects in 3GPP Next Generation Networks}
\label{sec:signaling}

To enable the proposed content-aware rate allocation frameworks in future wireless systems, particularly 6G, robust and scalable signaling mechanisms are required to facilitate the exchange of information between applications and the network. Table~\ref{tab:info_exchange} summarizes such required information exchange for the different proposed schemes. This section outlines several promising signaling options, many of which build upon enhancements introduced in 5G and are being extended in 3GPP and IETF standardization efforts.

\subsection{Protocol Signaling Options for Content-Awareness}
Several signaling mechanisms are being considered or standardized to support UX awareness in 5G and 6G networks:

\subsubsection{SCONE} 
The Standard Communication with Network Elements (SCONE) protocol, currently under development by the IETF, enables applications to directly exchange information with network elements in a standardized manner \cite{scone2025}. SCONE is designed to be transport-agnostic and compatible with various access technologies, including Wi-Fi, 5G, and 6G. It has been already proposed to use TRAIN (Transparent Rate Adaptation Indications for Networks) packets in the SCONE protocol to allow network elements to signal rate-limiting policies to endpoints in a transparent and advisory manner. These TRAIN packets can directly be used for the communication of assigned bitrate from RAN to AS/AC. These packets can be easily extended to carry QB curve or UX satisfaction factor from AS/AC to network elements (e.g., RAN).

\subsubsection{AQP}
Alternative QoS Profiles (AQP) is a mechanism introduced in 3GPP Release 16 to enhance the flexibility of QoS provisioning in mobile networks. Traditionally, a single QoS profile is associated with a PDU session, which limits the network's ability to adapt to varying application requirements and channel conditions. AQP addresses this limitation by allowing the Application Function (AF) to define multiple QoS parameter sets—each characterized by Guaranteed Flow Bit Rate (GFBR), Maximum Flow Bit Rate (MFBR), and other QoS parameters. These profiles are communicated to the Session Management Function (SMF), which forwards them to the RAN. The RAN then selects the most appropriate profile based on current network conditions and user-specific content complexity. This dynamic selection enables the network to optimize resource allocation not only based on channel conditions but also on the actual media content being consumed by users.

In the context of content-aware networking, AQP can serve as a powerful signaling tool to convey QB curves from the application to the network. Each QoS profile in the AQP set can represent a point on the UX-bitrate curve, mapping a specific bitrate to its corresponding QoE level (e.g., PSNR). By exposing this information to the RAN, the network becomes capable of making informed decisions that balance user satisfaction across multiple users in a cell. 
The use of AQP for UX signaling is further supported by ongoing 3GPP work on QoE measurement collection and management orchestration \cite{3gpp.28.307}.

\subsubsection{PDU-set metadata}
Packet Data Unit (PDU) sets are a concept introduced in 3GPP to represent a collection of packets that together carry a single media unit, such as a video frame \cite{gul2025pdu}. This abstraction is particularly useful for real-time media applications, where each frame must be delivered within strict latency bounds to ensure smooth playback and satisfactory UX. A PDU set may include metadata that describes the characteristics of the media unit it carries—such as its size, periodicity, jitter, and encoding complexity. This metadata can be embedded in application-layer headers (e.g., RTP Header Extensions, QUIC metadata, or MPEG-DASH descriptors) and extracted by network elements like the User Plane Function (UPF) or the Radio Access Network (RAN).
In the context of content-aware networking, PDU-set metadata serves as a lightweight and scalable mechanism to convey content complexity information (e.g., the QB curve) from the AS to the network.

\subsubsection{MAC-CEs}
MAC Control Elements (MAC-CEs) are compact signaling structures defined within the Medium Access Control (MAC) layer of 3GPP systems \cite{3gpp.38.321}, used to convey control information between the User Equipment (UE) and the Radio Access Network (RAN). These elements are designed to be lightweight and time-sensitive, making them ideal for real-time communication scenarios such as scheduling requests, buffer status reports, and power headroom indications. Each MAC-CE is identified by a unique Logical Channel ID (LCID) and follows a standardized format to ensure interoperability across devices and network implementations.

Some existing MAC-CEs may be reused to carry some of the information needed for our proposed schemes. For instance, the Recommended Bitrate (RBR) MAC-CE is designed to facilitate dynamic bitrate adaptation between the UE and the RAN. RBR MAC-CE enables the UE to communicate its desired bitrate to the network, which in turn responds with a recommended bitrate that reflects current network conditions and scheduling constraints. Hence, it can be used directly by RAN to convey its assigned bitrate (e.g., in the network-centered optimal schemes) to the UE/AC. However, while existing MAC-CEs support a range of control functions, they are not inherently designed to carry application-level or UX-centric information. To enable content-aware networking and support the proposed multi-user content diversity framework, new MAC-CEs could be defined to convey metrics such as the network-wide link price and the UX satisfaction factor. The link price, used in decentralized rate control algorithms, reflects the current level of network congestion and can guide UEs in adjusting their bitrate accordingly. Similarly, the UX satisfaction factor, used in content-aware PF scheduling scheme, can be reported by the UE to inform the RAN scheduler of its current quality state. Defining new MAC-CEs for these parameters would allow for low-latency, scalable signaling that integrates seamlessly with existing MAC procedures, while enabling the network to make more intelligent, user-centric resource allocation decisions.

\subsection{Protocol Signaling Considerations}
The practical deployment of content-aware signaling mechanisms in future wireless systems must address several integration and scalability challenges to ensure feasibility across diverse network environments and device ecosystems.
\begin{itemize}
    \item \textbf{Protocol Compatibility:} To minimize overhead and accelerate adoption, UX signaling should be embedded within existing protocol frameworks wherever possible.
    \item \textbf{Reporting Policies:} Reporting policies must be carefully designed to balance responsiveness with scalability. Frequent updates of UX information may improve adaptation accuracy but can also lead to excessive signaling overhead, particularly in dense deployments. Networks should support both periodic and event-driven reporting modes, where updates are triggered only when significant changes occur in bitrate demand or user experience metrics. Configurable thresholds and timers can help tailor these policies to specific application types and traffic profiles.
    \item \textbf{Security and Trust:} Trust is critical to ensure the integrity of UX-aware decision-making. Since UX feedback originates from application-layer entities, mechanisms must be put in place to validate the authenticity and reliability of the reported information. This may involve integrity checks, or cross-verification with network-side measurements. Without such safeguards, malicious or misconfigured applications could manipulate UX metrics to gain unfair resource advantages, undermining fairness and efficiency.
    \item \textbf{Standardization:} Continued collaboration with standardization bodies such as 3GPP and IETF is necessary to formalize the signaling formats, procedures, and semantics associated with content-aware networking. 
\end{itemize}

\section{Conclusions and Future Work}
\label{sec:conclusion}
In this paper, we introduced the concept of Multi-User Content Diversity as a foundational shift in how wireless networks allocate resources for immersive applications such as XR, cloud gaming, and real-time video streaming. Unlike traditional rate control mechanisms that rely solely on channel conditions, our proposed framework incorporates content complexity awareness to optimize UX across diverse media streams.
We developed optimal rate allocation algorithms that leverage application-provided content complexity information and real-time network metrics. These algorithms were shown to significantly improve key performance indicators such as UX capacity, fairness, and network utilization, achieving up to 67\% gains in UX capacity compared to conventional congestion control methods. Furthermore, our simulation results also demonstrate that even lightweight solutions, such as UX-aware PF scheduler, can yield substantial improvements in UX capacity, user satisfaction, and resource efficiency, with very little compromise in complexity. These findings underscore the potential of content-aware networking as a critical enabler for future 6G systems.

Looking ahead, several avenues remain open for exploration. For instance, the results of this paper assumed the perfect generation/estimation of QB curves at the application server. Practically, real-time RF curve estimation may be imperfect, and the impact of this imperfection on the proposed rate control algorithms needs to be studied. The application of the proposed algorithms in AI native networks, where the network resources are allocated using data-driven approaches, is also an interest line of future work. Additionally, the simulations in this paper assumed all UEs have similar traffic characteristics (real-time media). Handling  scenarios of the coexistence of UEs with different traffic types that require different UX modeling is an interesting venue and part of future work.

\bibliographystyle{IEEEtran}
\bibliography{refs.bib}

@inproceedings{nadas2024qoe,
  title={{To QoE or not to QoE}},
  author={N{\'a}das, Szilveszter and Ernstr{\"o}m, Lars and Szil{\'a}gyi, L{\'a}szl{\'o} and Patra, Gyanesh and Krylov, Dmitri and Lynam, Jonathan},
  booktitle={Proceedings of the 2024 Applied Networking Research Workshop},
  pages={38--44},
  year={2024}
}

@techreport{3gpp.38.835,
 author = {3GPP},
 institution = {{3rd Generation Partnership Project}},
 note = {Version 1.0.1},
 number = {38.835},
 title = {{Study on XR enhancements for NR}},
 type = {Technical Report},
 year = {2023}
}

@techreport{3gpp.38.901,
 author = {3GPP},
 institution = {{3rd Generation Partnership Project}},
 note = {Version 14.3.0},
 number = {38.901},
 title = {{Study on channel model for frequencies from 0.5 to 100 GHz}},
 type = {Technical Specification},
 year = {2018}
}

@techreport{3gpp.23.501,
 author = {3GPP},
 institution = {{3rd Generation Partnership Project}},
 note = {Version 19.0.0},
 number = {23.501},
 title = {{System architecture for the 5G System}},
 type = {Technical Specification},
 year = {2024}
}

@techreport{3gpp.28.307,
  title        = {{Telecommunication management; Quality of Experience (QoE) measurement collection Integration Reference Point (IRP); Requirements}},
  author       = {{3rd Generation Partnership Project (3GPP)}},
  institution  = {3GPP},
  type         = {Technical Specification},
  number       = {TS 28.307},
  version      = {18.0.0},
  year         = {2024}
}

@techreport{3gpp.38.321,
  title        = {{NR; Medium Access Control (MAC) protocol specification}},
  author       = {{3rd Generation Partnership Project (3GPP)}},
  institution  = {3GPP},
  type         = {Technical Specification},
  number       = {TS 38.321},
  version      = {18.6.0},
  year         = {2025}
}

@techreport{3gpp.22.870,
  title        = {{Study on 6G Use Cases and Service Requirements}},
  author       = {{3rd Generation Partnership Project (3GPP)}},
  institution  = {3GPP},
  type         = {Technical Report},
  number       = {TR 22.870},
  year         = {2025},
  note         = {Release 20}
}

@techreport{3gpp.26.928,
  title        = {{Extended Reality (XR) in 5G}},
  institution  = {3rd Generation Partnership Project (3GPP)},
  type         = {Technical Report},
  number       = {26.928},
  version      = {16.0.0},
  year         = {2020}
}

@article{maura2024experimenting,
  title={{Experimenting with Adaptive Bitrate Algorithms for Virtual Reality Streaming over Wi-Fi}},
  author={Maura, Ferran and Casasnovas, Miguel and Bellalta, Boris},
  journal={arXiv preprint arXiv:2407.15614},
  year={2024}
}

@inproceedings{korhonen2012peak,
  title={{Peak signal-to-noise ratio revisited: Is simple beautiful?}},
  author={Korhonen, Jari and You, Junyong},
  booktitle={2012 Fourth international workshop on quality of multimedia experience},
  pages={37--38},
  year={2012},
  organization={IEEE}
}

@article{briscoe2019implementing,
  title={{Implementing the’Prague Requirements’ for Low Latency Low Loss Scalable Throughput (L4S)}},
  author={Briscoe, Bob and others},
  journal={Netdev 0x13},
  year={2019}
}

@article{amiri2024application,
  title={{Application Awareness for Extended Reality Services: 5G-Advanced and Beyond}},
 author={Amiri, Abolfazl and others},
  journal={IEEE Communications Magazine},
  volume={62},
  number={8},
  pages={38--44},
  year={2024},
  publisher={IEEE}
}

@article{min2024perceptual,
  title={{Perceptual video quality assessment: A survey}},
  author={Min, Xiongkuo and Duan, Huiyu and Sun, Wei and Zhu, Yucheng and Zhai, Guangtao},
  journal={arXiv preprint arXiv:2402.03413},
  year={2024}
}

@inproceedings{yan2022qoe,
  title={{QoE-aware resource allocation for semantic communication networks}},
  author={Yan, Lei and Qin, Zhijin and Zhang, Rui and Li, Yongzhao and Li, Geoffrey Ye},
  booktitle={IEEE Global Communications Conference (GLOBECOM)},
  pages={3272--3277},
  year={2022},
  organization={IEEE}
}

@inproceedings{slivar2019qoe,
  title={{QoE-aware resource allocation for multiple cloud gaming users sharing a bottleneck link}},
  author={Slivar, Ivan and Skorin-Kapov, Lea and Suznjevic, Mirko},
  booktitle={2019 22nd conference on innovation in clouds, internet and networks and workshops (ICIN)},
  pages={118--123},
  year={2019},
  organization={IEEE}
}

@article{hande2023extended,
  title={{Extended reality over 5G—Standards evolution}},
  author={Hande, Prashanth and others},
  journal={IEEE Journal on Selected Areas in Communications},
  volume={41},
  number={6},
  pages={1757--1771},
  year={2023},
  publisher={IEEE}
}

@misc{rfc9330,
    series =    {Request for Comments},
    number =    9330,
    howpublished =  {RFC 9330},
    publisher = {RFC Editor},
    doi =       {10.17487/RFC9330},
    url =       {https://www.rfc-editor.org/info/rfc9330},
    author =    {Bob Briscoe and Koen De Schepper and Marcelo Bagnulo and Greg White},
    title =     {{Low Latency, Low Loss, and Scalable Throughput (L4S) Internet Service: Architecture}},
    pagetotal = 36,
    year =      2023,
    month =     jan,
}

@article{oliveira2020enhancement,
  title={{An enhancement of the bisection method average performance preserving minmax optimality}},
  author={Oliveira, Ivo and Takahashi, Ricardo},
  journal={ACM Transactions on Mathematical Software},
  volume={47},
  number={1},
  pages={1--24},
  year={2020},
  publisher={ACM New York, NY, USA}
}

@article{liebl2005radio,
  title={Radio link buffer management and scheduling for wireless video streaming},
  author={Liebl, G{\"u}nther and Jenkac, Hrvoje and Stockhammer, Thomas and Buchner, Christian},
  journal={Telecommunication Systems},
  volume={30},
  pages={255--277},
  year={2005},
  publisher={Springer}
}

@article{wang2006application,
  title={Application-oriented flow control: fundamentals, algorithms and fairness},
  author={Wang, Wei-Hua and Palaniswami, Marimuthu and Low, Steven H},
  journal={IEEE/ACM Transactions On Networking},
  volume={14},
  number={6},
  pages={1282--1291},
  year={2006},
  publisher={IEEE}
}

@article{zheng2024video,
  title={Video quality assessment: A comprehensive survey},
  author={Zheng, Qi and Fan, Yibo and Huang, Leilei and Zhu, Tianyu and Liu, Jiaming and Hao, Zhijian and Xing, Shuo and Chen, Chia-Ju and Min, Xiongkuo and Bovik, Alan C and others},
  journal={arXiv preprint arXiv:2412.04508},
  year={2024}
}

@article{korany2025ux,
  title={{UX-aware Rate Allocation for Real-Time Media}},
  author={Korany, Belal and Tinnakornsrisuphap, Peerapol and Kassir, Saadallah and Hande, Prashanth and Lee, Hyun Yong and Stockhammer, Thomas},
  journal={arXiv preprint arXiv:2505.04114},
  year={2025}
}

@inproceedings{rassool2017vmaf,
  title={{VMAF reproducibility: Validating a perceptual practical video quality metric}},
  author={Rassool, Reza},
  booktitle={2017 IEEE international symposium on broadband multimedia systems and broadcasting (BMSB)},
  pages={1--2},
  year={2017},
  organization={IEEE}
}

@techreport{3gpp.38.838,
  title        = {{Study on XR (Extended Reality) evaluations for NR}},
  institution  = {3rd Generation Partnership Project (3GPP)},
  number       = {TR 38.838},
  type         = {Technical Report},
  year         = {2021},
  note         = {Release 17},
}

@article{kelly1998rate,
  title={Rate control for communication networks: shadow prices, proportional fairness and stability},
  author={Kelly, Frank P and Maulloo, Aman K and Tan, David KH},
  journal={Journal of the Operational Research Society},
  volume={49},
  number={3},
  pages={237--252},
  year={1998},
  publisher={Springer}
}

@inproceedings{Tse2001Multiuser,
  author    = {David Tse},
  title     = {Multiuser Diversity in Wireless Networks: Smart Scheduling, Dumb Antennas and Epidemic Communication},
  booktitle = {IMA Wireless Workshop},
  year      = {2001},
  month     = {August}
}

@inproceedings{tu2005rate,
  title={Rate-distortion estimation for H. 264/AVC coders},
  author={Tu, Yu-Kuang and Yang, Jar-Ferr and Sun, Ming-Ting},
  booktitle={2005 IEEE International Conference on Multimedia and Expo},
  pages={4--pp},
  year={2005},
  organization={IEEE}
}

@misc{scone2025,
  author       = {Martin Thomson and Christian Huitema and Kazuho Oku and Matt Joras and Marcus Ihlar},
  title        = {{Standard Communication with Network Elements (SCONE) Protocol}},
  howpublished = {Internet-Draft},
  institution  = {IETF},
  year         = {2025},
  month        = {July},
  url          = {https://www.ietf.org/archive/id/draft-ietf-scone-protocol-01.html}
}

@article{gul2025pdu,
  title={{PDU Set based QoS Handling in 3GPP: Release 18 Overview and Future Directions}},
  author={G{\"u}l, Serhan and Ahsan, Saba and Paris, Stefano and Curcio, Igor DD},
  journal={IEEE Communications Standards Magazine},
  year={2025},
  publisher={IEEE}
}

@techreport{ericsson_mobility_2018,
  title        = {Ericsson Mobility Report – November 2018},
  author       = {{Ericsson}},
  institution  = {Ericsson},
  year         = {2018},
  month        = {November},
  url          = {https://www.ericsson.com/en/reports-and-papers/mobility-report/reports},
  note         = {Available online}
}

@inproceedings{Altman2008AlphaFair,
  author    = {Eitan Altman and Konstantin Avrachenkov and Andrey Garnaev},
  title     = {Generalized $\alpha$-Fair Resource Allocation in Wireless Networks},
  booktitle = {Proceedings of the 47th IEEE Conference on Decision and Control (CDC)},
  year      = {2008},
  pages     = {2414--2419}
}

@article{rezaei2008joint,
  title={{Joint video coding and statistical multiplexing for broadcasting over DVB-H channels}},
  author={Rezaei, Mehdi and Bouazizi, Imed and Gabbouj, Moncef},
  journal={IEEE Transactions on Multimedia},
  volume={10},
  number={8},
  pages={1455--1464},
  year={2008},
  publisher={IEEE}
}

\end{document}